\documentstyle[11pt,paspconf,epsf]{article}

\newcommand\lta{\mathrel{\rlap{\lower 3pt\hbox{$\mathchar"218$}}
     \raise 2.0pt\hbox{$\mathchar"13C$}}}
\newcommand\gta{\mathrel{\rlap{\lower 3pt\hbox{$\mathchar"218$}}
     \raise 2.0pt\hbox{$\mathchar"13E$}}}
\newcommand\kms{km~s$^{-1}$}
\newcommand\kmsM{km~s$^{-1}\,$Mpc}

\newcommand\etal{{et~al.}} 
\newcommand\sigv{\ifmmode \sigma_v\else$\sigma_v$\fi}
\newcommand\bi{\ifmmode \beta_{I}\else$\beta_I$\fi}
\newcommand\bo{\ifmmode \beta_{O}\else$\beta_O$\fi}
\newcommand\czd{$c{z}$--$d$}
\def\iras{{\it IRAS}}

\begin{document}

\title{Comparing the SBF Survey Velocity Field with the
Gravity Field from Redshift Surveys}

\author{\sc John P. Blakeslee}
\vskip -2pt
\affil{Palomar Observatory, Caltech, MS 105-24, Pasadena, CA 91125}
\author{\sc Marc Davis}
\vskip -2pt
\affil{Dept.~of Astronomy, University of California,
Berkeley, CA~94720}
\author{\sc John L. Tonry}
\vskip -2pt
\affil{Institute for Astronomy,
University of Hawaii, Honolulu, HI 96822}
\author{\sc Edward A. Ajhar}
\vskip -2pt
\affil{Kitt Peak National Observatory, 
P.\,O. Box 26732, Tucson, AZ 85726}
\author{\sc Alan Dressler}
\vskip -2pt
\affil{Carnegie Observatories, 813 Santa Barbara St.,
Pasadena, CA 91101}


\begin{abstract}
We compare the predicted local peculiar velocity field from the \iras\
1.2~Jy flux-limited redshift survey and the Optical Redshift Survey
(ORS) to the measured peculiar velocities from the recently completed
SBF Survey of Galaxy Distances. The analysis produces a value of
$\beta{\,=\,}\Omega^{0.6}/b$ for the redshift surveys, where $b$ 
is the linear biasing factor, and a tie to the Hubble flow, 
i.e., a value of $H_0$, for the SBF Survey. 
There is covariance between these parameters, but we find 
good fits with $H_0 \approx 74$ \kmsM\ for the SBF distances,
$\bi \approx 0.44$ for the \iras\ survey predictions, and 
$\bo \approx 0.3$ for the ORS.
The small-scale velocity error $\sigv\sim200$ \kms\ is
similar to, though slightly larger than, the value obtained
in our parametric flow modeling with SBF.
\end{abstract}


\section{The SBF Survey}

The Surface Brightness Fluctuation (SBF) method of estimating early-type
galaxy distances has been around for over a decade (Tonry \& Schneider 1988).
Blakeslee, Ajhar, \& Tonry (1999) have recently reviewed 
the SBF method, its applications, and calibration.
Tonry, Ajhar, \& Luppino (1990) made the first application
in the Virgo cluster and calibrated it using theoretical
isochrones for the color dependence combined with the Cepheid distance to
M31 for the zero~point.  Tonry (1991) applied the method to Fornax
galaxies and made the first fully empirical calibration, 
differing substantially from the earlier theoretical one.
The situation is now much improved, with the best theoretical
models (Worthey 1994) agreeing well with the latest empirical calibration.

The $I$-band SBF Survey of Galaxy distances began in earnest in
the early 1990's using the 2.4\,m telescope at MDM Observatory on
Kitt Peak in the Northern hemisphere and the 2.5\,m at Las Campanas
Observatory in the South.  The Survey includes distances to over
330 galaxies reaching out to $c{z} \sim 4500$ \kms.
Tonry \etal\ (1997; hereafter SBF-I) describe the SBF Survey sample in detail.
The breakdown by morphology is 55\% ellipticals, 40\% lenticulars,
and 5\% spirals.  The median distance error is about 0.2~mag,
several times larger than our estimate of the intrinsic scatter
in the method.  This is due to the compromises involved in conducting
a large survey with an observationally demanding method on small
telescopes; thus, most individual galaxy distances could be
improved with reobservation in more favorable conditions.

Tonry \etal\ (1999, hereafter SBF-II) investigate the large-scale flow
field within and around the Local Supercluster using extensive
parametric modeling.  This modeling is summarized by Tonry \etal\ and
Dressler \etal\ in the present volume.  Also in this volume are
SBF-related works by Pahre \etal\ calibrating the $K$-band fundamental
plane with SBF Survey distances and by Liu
\etal\ reporting $K$-band SBF measurements in the Fornax and Coma
clusters.  Here we present a preliminary comparison of the SBF survey
peculiar velocities with expectations from the density field as probed
by redshift surveys. 
More details on many aspects of this work
are given by Blakeslee \etal\ (2000).

\section{Comparing to the Density Field}

\begin{figure}
\plotone{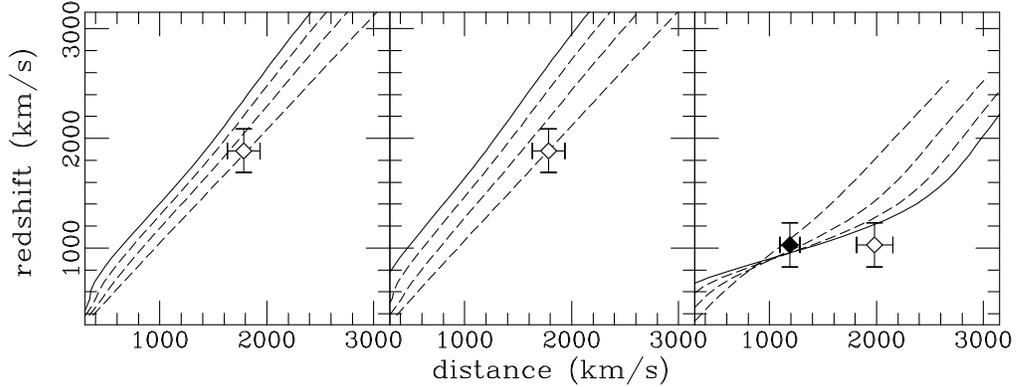}
\caption{\small 
Predicted \czd\ relations for some
individual galaxies.  The points are placed according to
their observed velocity and SBF distance; the curves are
predictions for $\beta=0.1,0.4,0.7,1.0$, with 
$\beta{\,=\,}1$ being the solid curve.  The overall 
distance scale (i.e., $H_0$) is a free parameter; thus
for each $\beta$, the $\chi^2$ analysis
allows the data points to slide uniformly in the horizontal direction.
The left and center panels exhibit the same galaxy (NGC\,821) with
\czd\ curves from the \iras\ survey (left) and the ORS (center).
The ORS curves fan out more at high $\beta$.  They also ``bend'' more
in the presence of mass concentrations, but the right panel shows that
even the $\beta{\,=\,}1$ ORS predictions from the spherical
harmonic method cannot accommodate a galaxy behind Virgo with a large 
infall velocity (open point), i.e., they cannot reproduce multivalued zones.
Note that a galaxy near the center of Virgo (solid point) helps little
in distinguishing between the $\beta$~curves, which cross at this point.
\label{fig:czd}}
\end{figure}

We use the method of Nusser \& Davis (1994; see the
description by Davis in this volume) to perform a spherical 
harmonic solution of the gravity field from the observed
galaxy distribution of both the \iras\ 1.2~Jy flux-limited
redshift survey (Strauss \etal\ 1992; Fisher \etal\ 1995) and 
the Optical Redshift Survey (ORS) (Santiago \etal\ 1995).
We assume linear biasing so that the fluctuations in the galaxy number density 
field are proportional to the mass density fluctuations, i.e.,
$\delta_g = b\,\delta_m$, where $b$ is the bias factor,
and use linear gravitational instability theory 
(e.g., Peebles 1993) so that the predicted peculiar velocities
are determined by $\beta{\,\equiv\,}\Omega^{0.6}/b$.
We then compute the distance-redshift relation in the
direction of each sample galaxy as a function of $\beta$.

The comparisons are done with the same subset of SBF survey
galaxies as used in SBF-II: galaxies with good quality data,
$(V{-}I)_0 > 0.9$ so that the color calibration applies, and not
extreme in their peculiar velocities (e.g., no Cen-45);  
we also omit Local Group members.  The sample is then 280 galaxies.
We compare the predicted \czd\ relations to the observations
using a simple $\chi^2$ minimization approach, as adopted by
Riess \etal\ (1997) in the comparison to the Type~Ia supernova (SNIa)
distances. Figure~\ref{fig:czd} gives an illustration.
Unlike SNIa distances however, the SBF distances have no secure external tie
to the far-field Hubble flow, so we allow the overall scale of
the distances (in km/s) to be a free parameter.  The best-fit 
scale then yields a value for $H_0$ when combined with the SBF tie
to the Cepheids, which is uncertain at the $\pm0.1$~mag level (SBF-II).
Thus, the 15 or more free parameters of SBF-II are here replaced
by just two parameters: $H_0$ and $\beta$.

Our sample consists mainly of early-type galaxies in groups, with
the dominant groups being the Virgo and Fornax clusters.
We use a constant small-scale velocity error $\sigv=200$ \kms\
and deal with group/cluster virial dispersions by 
using group-averaged redshifts for the galaxies. The group
definitions are from SBF-I; more than a third of the galaxies
are not grouped.  This approach is unlike that of SBF-II,
which used only individual galaxy velocities but a variable \sigv.
Blakeslee \etal\ (2000) explore several approaches
in doing the comparison to the gravity field,
including one with no grouping but a variable \sigv,
and find similar results to what we report below.
In addition, they find negligibly different results when
$\sigv=150$ \kms\ is used instead of 200 \kms.

\begin{figure}
\plottwo{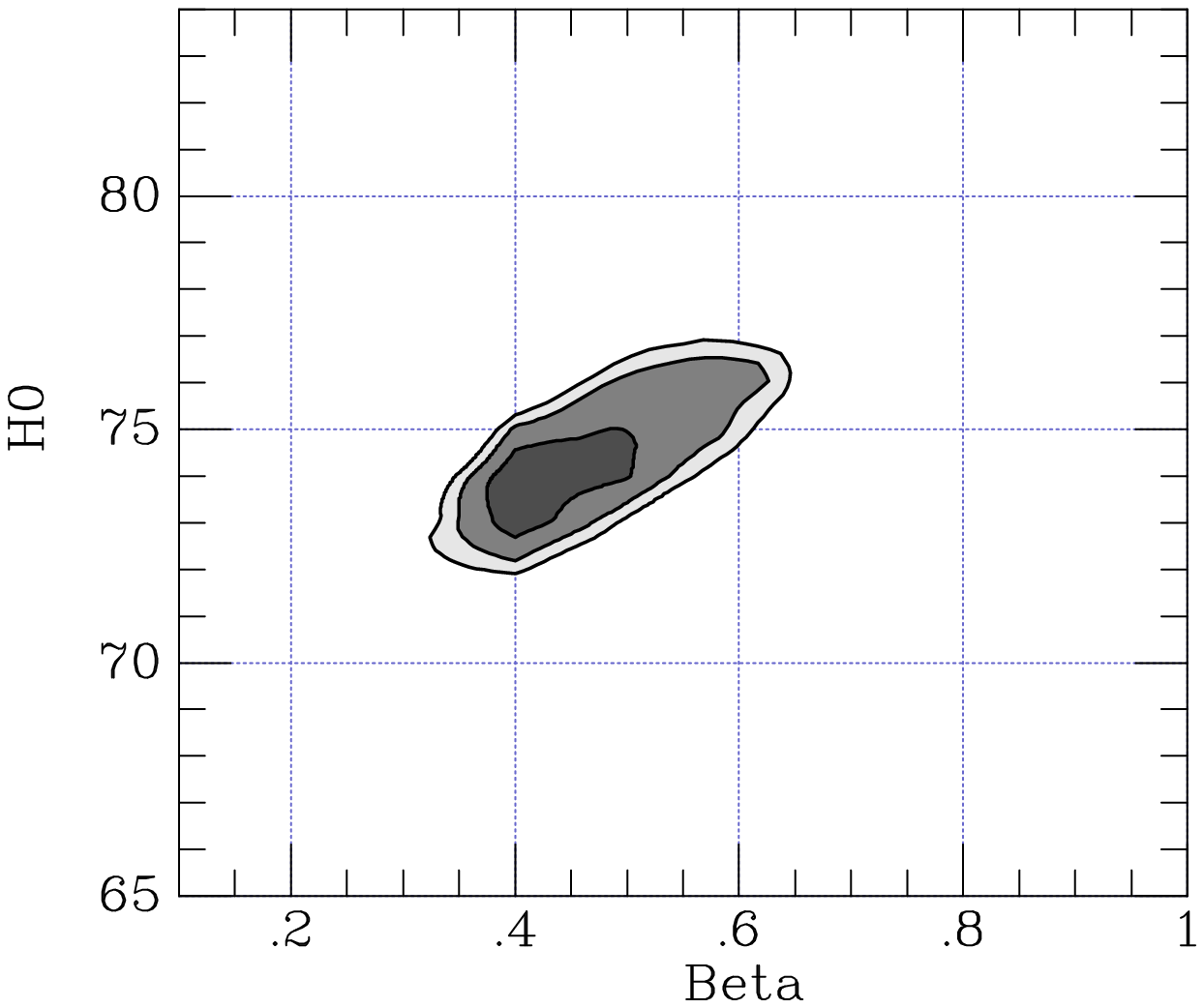}{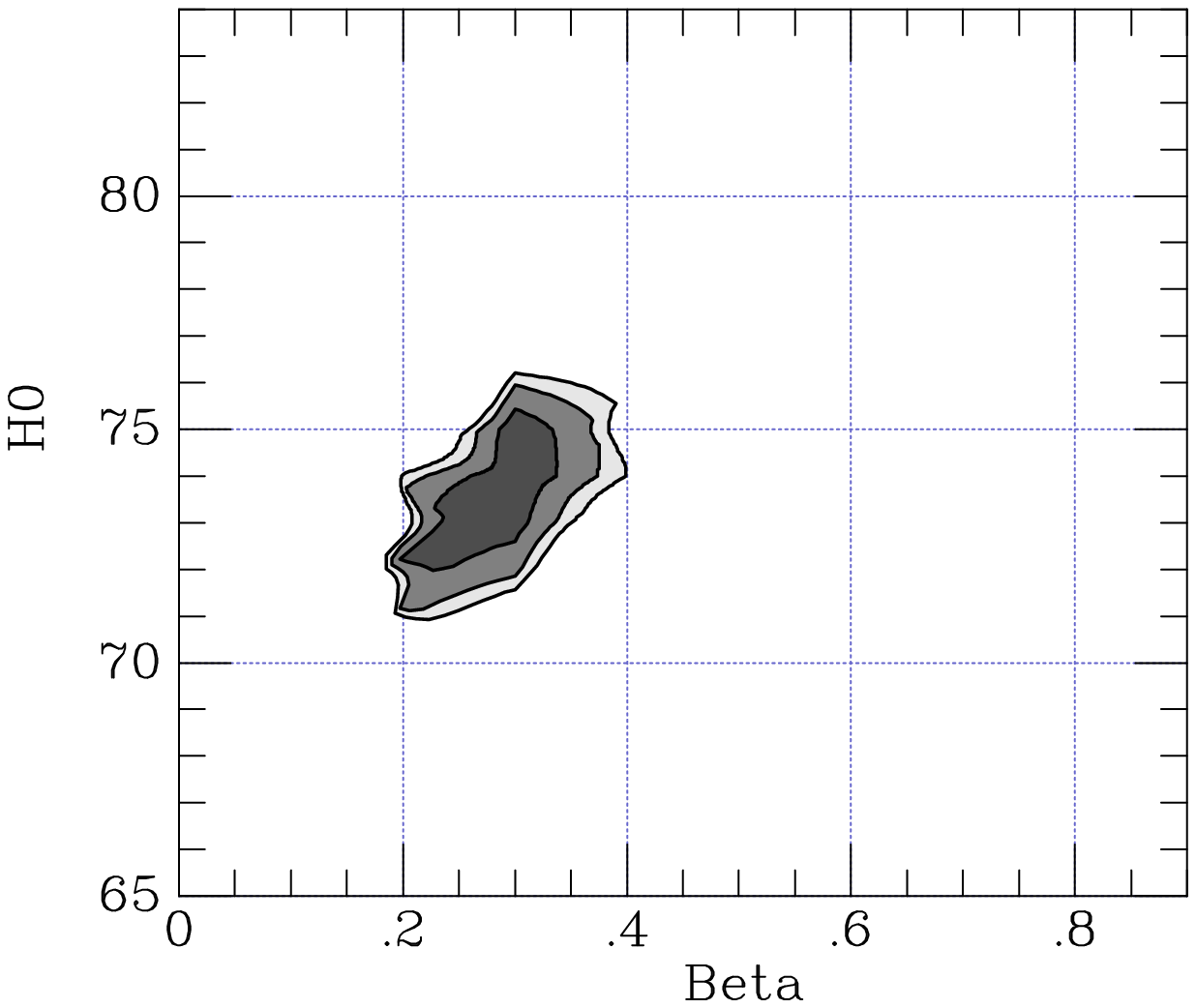}
\caption{\small The 68\%, 90\%, and 95\% joint probability contours 
on $H_0$ and~$\beta$ for the comparison between the SBF peculiar velocities
and the predictions from the \iras\ 1.2~Jy redshift survey (left)
and the ORS (right). The irregular, ragged contours are due to
the coarseness of the grids.
\label{fig:conts}}
\end{figure}

\section{Results}

\begin{figure}
\plotone{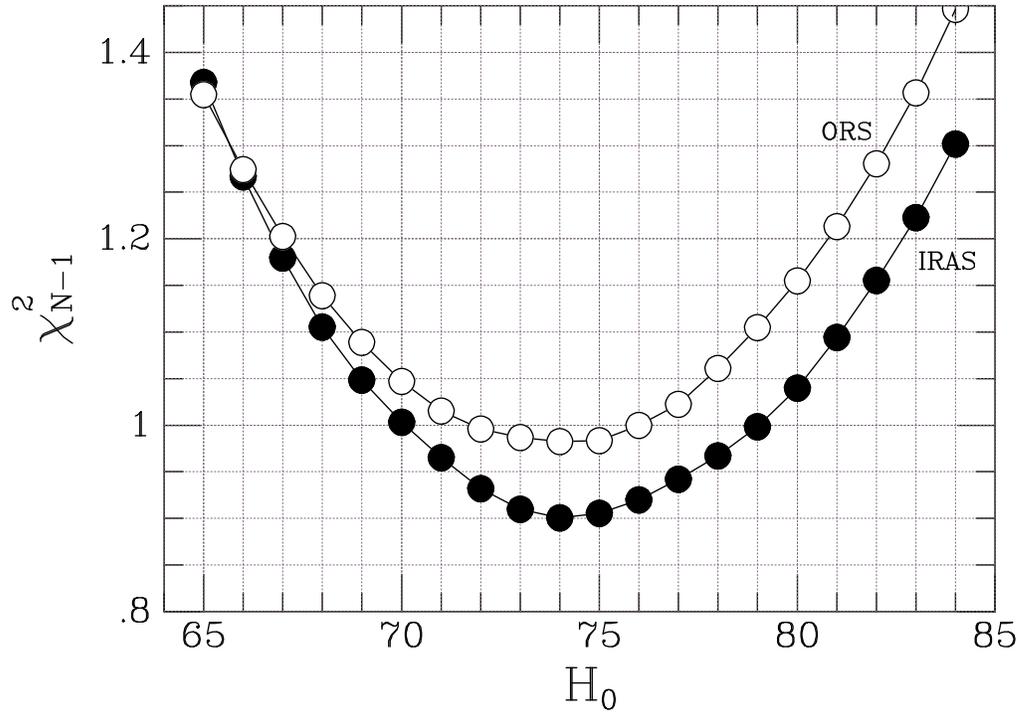}
\caption{\small Reduced $\chi^2$ as a function of $H_0$ for
$\bi = 0.45$ and $\bo=0.30$.  The minimum is reached near
$H_0 = 74$ or 74.5 \kmsM.
\label{fig:chiHo}}
\end{figure}

Figure~\ref{fig:conts} displays the joint probability contours on 
$H_0$ and $\beta$ derived from the $\chi^2$ analysis of the
\iras/SBF and ORS/SBF comparisons.  The calculations are done 
in $\beta$~steps of 0.1 and $H_0$ steps of 1\,\kmsM.
Both redshift surveys call for $H_0\approx74$ \kmsM.
For $\beta$, the \iras\ comparison gives $\bi=0.4$--0.5, while
the ORS prefers $\bo\approx0.3$.  Adopting the best $\beta$
models for each comparison, Figure~\ref{fig:chiHo} shows the
reduced $\chi^2$ for 279 degrees of freedom
plotted against $H_0$ (we averaged the
$\bi=0.4$ and $\bi=0.5$ predictions for \iras).
As there can only be one $H_0$, and the \iras\ and ORS surveys
are not wholly independent, it is perhaps not surprising
that they give consistent results.  Interestingly, the
preferred $H_0$ splits the difference between the ``SBF $H_0$'' 
values proffered by SBF-II and Ferrarese \etal\ (1999).

\begin{figure}
\plotone{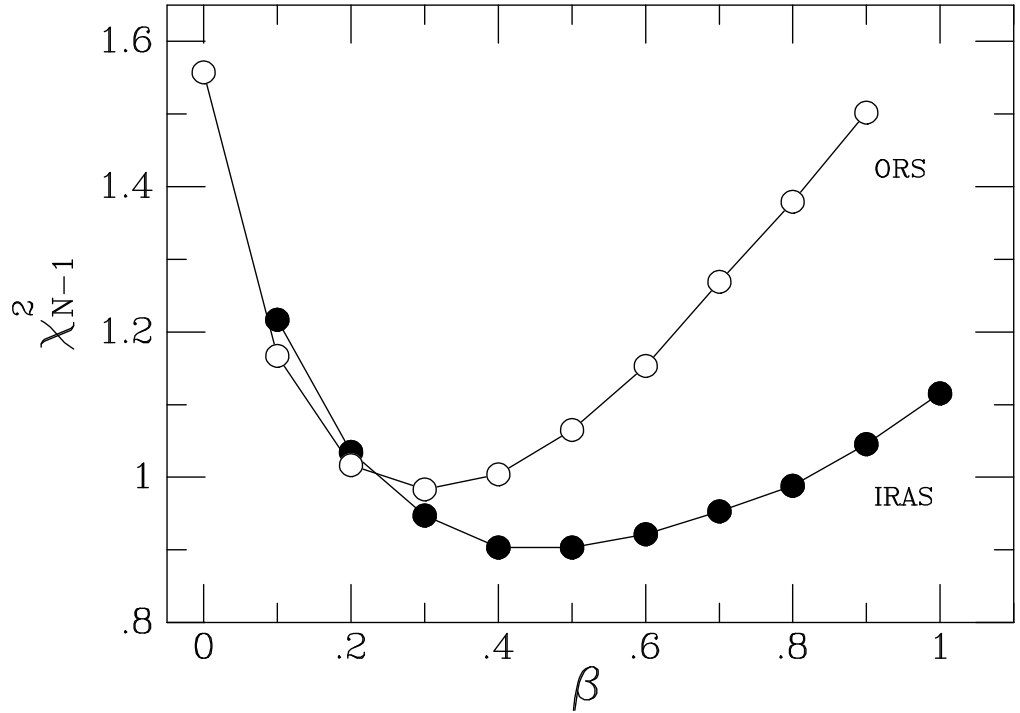}
\caption{\small Reduced $\chi^2$ as a function of $\beta$ for
$H_0 = 74.5$ \kmsM.
\label{fig:chibeta}}
\end{figure}

The reduced $\chi^2$ as a function of $\beta$ for a fixed $H_0$
is shown in Figure~\ref{fig:chibeta}.  We find $\bi=0.44\pm0.08$
for \iras\ and $\bo=0.30\pm0.06$ for the ORS with the adopted $H_0$.
We see that $\sigv=200$ \kms\ gives $\chi^2_\nu{\,=\,}1$ for the
ORS comparison, and 0.9 for the \iras\ comparison; the 
latter would have $\chi^2_\nu{\,=\,}1$ for $\sigv=180$ \kms,
very similar to the value of $\sigv$ found in the SBF-II parametric
modeling.  The derived $\beta$'s are not sensitive to the adopted
$\sigv$, but they are to $H_0$ because of the covariance seen in 
Figure~\ref{fig:conts}.
Had we adopted an $H_0$ 5\% larger, the best-fit $\beta$'s would
increase by $\sim\,$30\% with still reasonable values of $\chi^2$.

\begin{figure}
\plotone{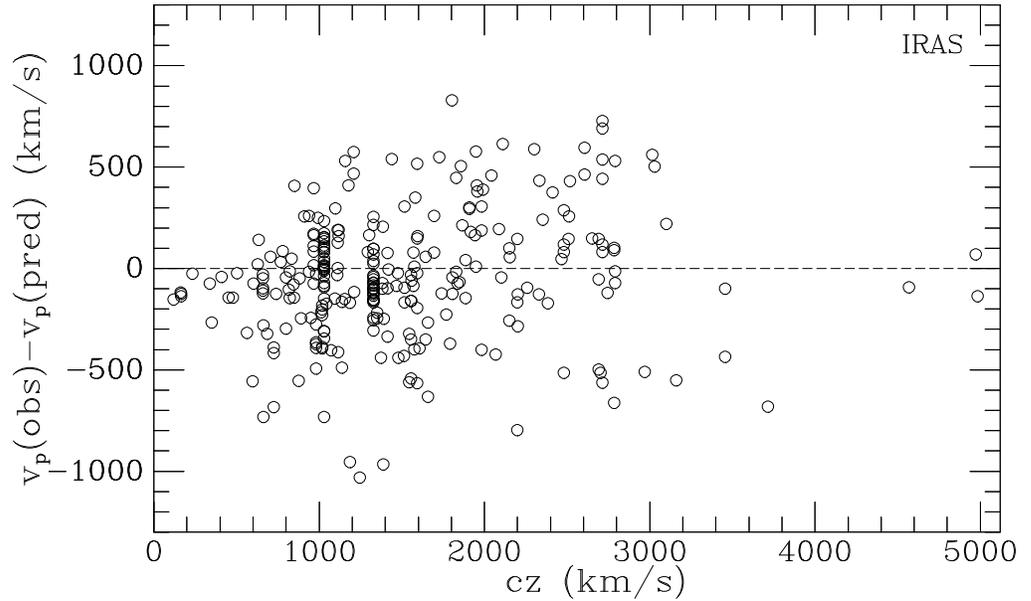}
\caption{\small The residuals (observed minus predicted peculiar
velocity) for the IRAS 1.2~Jy survey/SBF Survey comparison with the
best-fitting combination of $H_0$ and \bi.  The three most distant
points exhibit small scatter because they are high quality HST measurements.
\label{fig:resids}}
\end{figure}

We also note that the $\beta$'s are insensitive to our treatment
of the clusters.  We experimented by throwing out all galaxies
conceivably near the triple-valued zones of Virgo, Fornax, and
Ursa Major, 40\% of the sample.
Remarkably, the results differ only negligibly, but $\chi^2_\nu$
increases by about 0.15, and the values of \sigv\ giving $\chi^2_\nu$
of unity increase by 10\%.  The \iras\ residual plot for the 
full comparison shown in Figure~\ref{fig:resids}
demonstrates why.  The Virgo and Fornax clusters near $c{z}$ of
1000 and 1400 \kms, respectively, lie close to the zero-difference
line with fairly small scatter because they have had their
virial dispersions effectively removed, unlike most of the
rest of the galaxies.  However, for the reason illustrated in
Figure~\ref{fig:czd}, the clusters do not drive the fit.

\begin{figure}
\plotfiddle{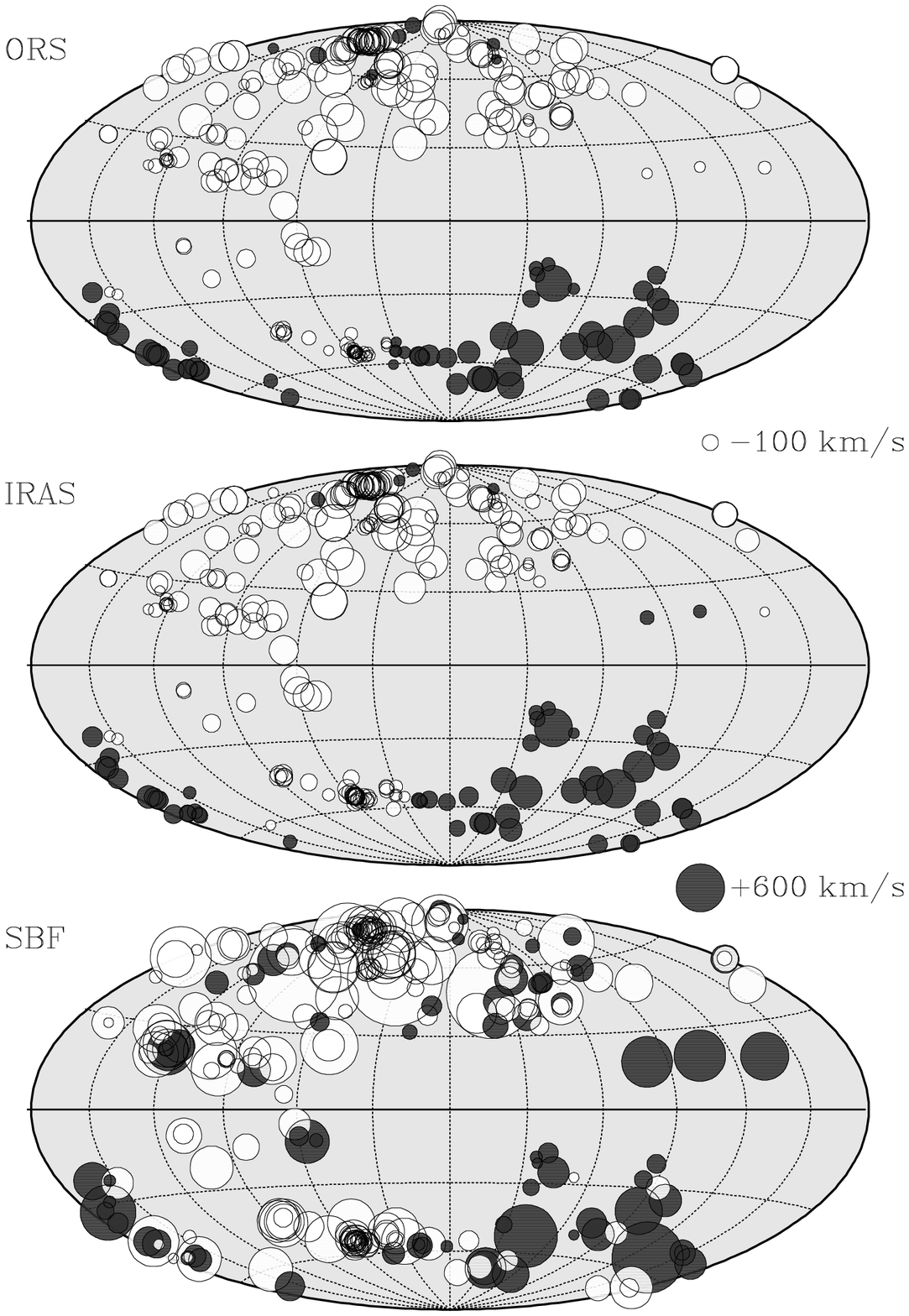}{6.7in}{0}{80}{80}{-266}{-77}
\caption{\small 
Predicted peculiar velocities from the ORS for $\bo=0.3$ (top) 
and \iras\ for $\bi=0.4$ (middle) are compared to observed peculiar
velocities from the SBF Survey (bottom) with $H_0=74$ \kmsM.
Velocities are in the Local Group frame and coordinates are Galactic.
Light/dark circles show negative/positive peculiar velocities,
coded in size as shown.
Unlike the predictions, the observations have no smoothing
applied and contain both distance and velocity errors,
causing the greater range in symbol size.
\label{fig:aitoffs}}
\end{figure}

Figure~\ref{fig:aitoffs} shows how the peculiar velocity predictions
and observations look on the sky.  The predictions should resemble
a noiseless, smoothed version of the observations.
Clearly there is general agreement on the most prominent feature,
the dipole motion seen in the Local Group frame. 
The observed large negative velocities near
$(l,b)\sim(283^\circ,+74^\circ)$ indicate rapidly infalling
Virgo background galaxies, which as noted in Figure~\ref{fig:czd},
the fit cannot reproduce.
This and other issues must be dealt with in future analyses.

\section{Summary and Future Prospects}

We have presented preliminary results from an initial comparison of
the SBF Survey peculiar velocities with predictions based on the
density field of the \iras\ and ORS redshift surveys.  The comparison
simultaneously yields $\beta$ for the density field and a tie to the
Hubble flow for SBF, i.e., $H_0$.  The resulting 
$H_0\approx74$ is between the two other recent estimates with SBF.  
For the \iras\ comparison we find $\bi\approx0.44$, consistent
with other recent results populating the 0.4--0.6 range
from ``velocity-velocity''
comparisons using Tully-Fisher or SNIa distances
(e.g., Schlegel 1995; Davis \etal\ 1996; Willick \etal\ 1997;
Riess \etal\ 1997; da\,Costa \etal\ 1998; Willick \& Strauss 1998).
Our value of $\bo\approx0.30$ is the same as that of Riess \etal,
and consistent with the expectation $\bo/\bi\sim0.7$ from
Baker \etal\ (1998).  The numbers change only slightly
with different treatments of the galaxy clusters and
higher resolution computations (Blakeslee \etal\ 2000).
Our results thus reinforce the
``factor-of-two discrepancy'' with the high $\beta$'s obtained in
``density-density'' comparisons (e.g., Sigad \etal\ 1998).
One explanation is a scale-dependent biasing (e.g., Dekel, this volume).

We plan in the near future to pursue comparisons using methods
that can deal with multivalued redshift zones directly, such 
VELMOD (Willick \etal\ 1997) and take advantage of SBF's 
potential for probing small, nonlinear scales.
Additionally, we continue to work towards an independent
far-field tie to the Hubble flow by measuring SBF distances
to SNIa galaxies, through calibration of the $K$-band 
fundamental plane Hubble diagram (Pahre \etal, this volume),
and by pushing out to $c{z}\sim10,000$ \kms\ directly using
SBF measurements from space.  This will remove the substantial
systematic uncertainty in $\beta$ due to the covariance with $H_0$.
Finally, we also plan to use SBF data for a ``density-density''
measurement of $\beta$ and to explore the nature of the biasing.

\acknowledgments
JPB thanks the Sherman Fairchild Foundation for support.
The SBF Survey was supported by NSF grant AST9401519.



\begin{references}
{ 
\reference
Baker, J. E., Davis, M., Strauss, M. A., Lahav, O., Santiago, B. X.
	1998, \apj, 508, 6
\reference
Blakeslee, J. P., Ajhar, E. A., \& Tonry, J. L. 1999, 
  in Post-Hipparcos Cosmic Candles, eds.\ A. Heck \& F. Caputo
  (Boston: Kluwer Academic),~181
\reference
Blakeslee, J. P., Davis, M., Tonry, J. L., Dressler, A., \& Ajhar, E. A.
   2000, \apj, in press
\reference
da\,Costa, L. N., Nusser, A., Freudling, W., Giovanelli, R., 
	Haynes, M. P., Salzer, J. J., \& Wegner, G. 1998, \mnras, 299, 425
\reference
Davis, M., Nusser, A. \& Willick, J. A. 1996, \apj, 473, 22
\reference
Ferrarese, L., \etal\ ($H_0$ Key Project) 1999, \apj, in press
\reference
Fisher, K. B., Huchra, J. P., Strauss, M. A., Davis, M., Yahil, A., \& 
   Schlegel, D.  1995, \apjs, 100,~69
\reference
Nusser, A. \& Davis, M. 1994, \apj, 421, L1
\reference
Peebles, P.J.E. 1993, Principles of Physical Cosmology
   (Princeton Univ.~Press)
\reference
Riess, A. G., Davis, M., Baker, J., \& Kirshner, R. P. 1997,
   \apj, 488, L1
\reference
Santiago, B. X., Strauss, M. A., Lahav, O., Davis, M., Dressler, A.,
	\& Huchra, J. P. 1995, \apj, 446, 457
\reference
Schlegel, D. J. 1995, Ph.D. Thesis, Univ. of California, Berkeley
\reference
Sigad, Y., Eldar, A., Dekel, A., Strauss, M. A., \& Yahil, A. 1998,
	\apj, 495, 516
\reference
Strauss, M. A., Huchra, J. P., Davis, M., Yahil, A., Fisher, K. B.,
\& Tonry, J. 1992, \apjs, 83, 29
\reference
Tonry, J. L. 1991, \apj, 373, L1.
\reference
Tonry, J. L., Ajhar, E. A., \& 	Luppino, G. A. 1990, \aj, 100, 1416
\reference
Tonry, J. L., Blakeslee, J. P.,
  Ajhar, E. A., \& Dressler, A. 1997, \apj, 475, 399 (SBF-I)
\reference
Tonry, J. L., Blakeslee, J. P.,
   Ajhar, E. A., \& Dressler, A. 1999, \apj, in press (SBF-II)
\reference
Tonry, J. L. \& Schneider, D. P. 1988, \aj, 96, 807
\reference
Willick, J. A., Strauss, M. A., Dekel, A., \& Kolatt, T. 1997, \apj, 486, 629
\reference
Willick, J. A. \& Strauss, M. A. 1998, \apj, 507, 64
}
\end{references}
\end{document}